\documentclass{aa}
\usepackage{graphicx}
\usepackage{natbib}
\usepackage{multirow}
\usepackage{txfonts}
\usepackage{hyperref}
\usepackage[small]{caption}
\bibpunct{(}{)}{;}{a}{}{,}
\hypersetup{colorlinks=true, citecolor=red, filecolor=blue,urlcolor=blue,
linkcolor=blue}

\begin{document}

\def\5921{COSMOS~5921+0638}
\title{COSMOS~5921+0638: Characterization and analysis of a new strong
gravitationally lensed AGN\thanks{Based on observations made with ESO Telescopes
at Paranal Observatory programme IDs: 077.A-0473(A) and 175.A-0839(B,D) and with
 NASA/ESA Hubble Space Telescope, obtained from the data archive at the Space
Telescope Institute. STScI is operated by the association of Universities for
Research in Astronomy, Inc. under the NASA contract  NAS 5-26555; also based on
data
collected at: the Subaru Telescope, which is operated by the National
Astronomical Observatory of Japan; the European Southern Observatory
under Large Program 175.A-0839, Chile; Kitt Peak National Observatory,
Cerro Tololo Inter-American Observatory, and the National Optical
Astronomy Observatory, which are operated by the Association of
Universities for Research in Astronomy, Inc. (AURA) under cooperative
agreement with the National Science Foundation; and the
Canada-France-Hawaii Telescope operated by the National Research
Council of Canada, the Centre National de la Recherche Scientifique de
France and the University of Hawaii.}}
\subtitle{}
\author{T. Anguita\inst{1}, C. Faure\inst{2}, J.-P. Kneib\inst{3}, J.
Wambsganss\inst{1}, C. Knobel\inst{4}, A. M. Koekemoer\inst{5} and M. Limousin\inst{3,6}}
\institute{Astronomisches Rechen-Institut, Zentrum f\"{u}r Astronomie der
Universit\"{a}t Heidelberg, M\"{o}nchhofstr. 12-14, D-69120, Heidelberg, Germany\\
e-mail: \texttt{tanguita@ari.uni-heidelberg.de}
\and
Laboratoire d'Astrophysique, Ecole Polytechnique F\'ed\'erale de Lausanne
(EPFL), Observatoire de Sauverny, 1290 Versoix, Switzerland
\and
 Laboratoire d'Astrophysique de Marseille, UMR\,6610, CNRS-Universit\'e de Provence, 38 rue Fr\'ed\'eric Joliot-Curie, 13\,388 Marseille Cedex 13, France
\and
Institute of Astronomy, Swiss Federal Institute of Technology (ETH Honggerberg),
CH-8093, Zurich, Switzerland
\and
Space Telescope Science Institute, 3700 San Martin Drive, Baltimore, MD 21218
\and
Dark Cosmology Centre, Niels Bohr Institute, University of Copenhagen, Juliane Maries Vej 30, 2100 Copenhagen, Denmark}
\date{Received 17 March 2009 / Accepted 23 July 2009}

\authorrunning{T. Anguita et al.}
\titlerunning{\5921:  Characterization and analysis of a new strong
gravitationally lensed AGN}

 \abstract{Strong lens candidates have been newly identified within the COSMOS field. We present VLT/FORS1 spectroscopic follow-up observations and HST/WFPC2 imaging of the system \5921, which exhibits quadruply lensed images and a perfect Einstein ring.}
{We investigate the nature of \5921\ by studying its photometric, spectroscopic and physical properties.}
{By analyzing our VLT/FORS1 spectroscopy and Subaru/CFHT/HST imaging of \5921, we completed both an environmental analysis and detailed analytical and grid-based mass modeling to determine it properties.}
{We measured the redshifts of the lensing galaxy in COSMOS 5921+0638
(z$_l$=0.551$\pm$0.001) and 9 additional galaxies in the field (5 of them at
z$\sim$0.35). The redshift of the lensed source was inferred by identifying a candidate Ly$\alpha$
line at z$_s$=3.14$\pm$0.05. The mass modeling reveals the requirement of a small
external shear ($\gamma$=0.038), which is slightly larger than the lensing
contribution expected by galaxy groups along the line-of-sight obtained from the zCOSMOS
optical group catalog ($\kappa_{groups}$$\sim$0.01 and
$\gamma_{groups}$$\sim$0.005). The estimated time-delays between the different
images are of the order of hours to half a week and the total magnification of the
background source is $\mu$$\approx$150. The measured mass-to-light ratio of the lensing galaxy within
the Einstein ring is $M/L_B$$\approx$8.5$\pm$1.6.
Anomalies are observed between the measured and expected flux ratios of the
images of the background AGN.}
{Our analysis indicates that the ring and point-like structures in \5921\ consist of a lensed high redshift galaxy hosting a low luminosity AGN (LLAGN). The observed flux ratio anomalies are probably due to microlensing by stars in the lensing
galaxy and/or a combination of static phenomena. Multi-epoch, multi-band space-based
observations would allow us to differentiate between the possible causes of these anomalies, since static and/or dynamic variations could be
identified. Because of its short time-delays and the possibility of microlensing, COSMOS 5921+0638 is a promising laboratory for future
studies of LLAGNs.}
\keywords{galaxies: individual: \5921 -- galaxies: quasars: individual: \5921 -- gravitational lensing -- cosmology: observations}

\maketitle

\section{Introduction}
\bigskip

Gravitationally lensed quasars are powerful observational tools in cosmology. The intrinsic variability of AGNs, provides a way to measure the time-delays between multiple quasar images \citep[an updated list of lensed quasars' time-delays was established by][]{oguri07}. By coupling these measurements with theoretical models for the lensing potential, the Hubble constant (H$_0$) can be derived \citep{refsdal64}. They allow us to investigate the amount and distribution of matter in lensing galaxies regardless of whether the matter is luminous or not. This property ensures that lensed quasars are ideal laboratories for probing for the dark matter distribution and dark matter substructures in and around lensing galaxies \citep{mao98,chiba02,yonehara03,kochanekdalal04,keeton08,trott08}. Furthermore, the natural magnification of background sources offers a means of studying the properties of objects in the high redshift universe. In particular, if lensed AGN images are affected by brightness fluctuations produced by individual stars in lensing galaxies (microlensing), the inner structure (i.e., milliparsec scales) of the AGN accretion disks can be probed \citep{wambsganss90,yonehara99,kochanek04,anguita08a,eigenbrod08}.

Strong lens samples \citep[e.g.,][]{myers01,fassnacht04,cabanac07,bolton08,inada08,faure08a} allow us to study in a statistical way the properties of lensing galaxies \citep{treu08} and the possible contribution of the environment to the formation of the lens \citep{oguri05,dobke07,faure08b}. To fully understand the properties of a lens sample, we need measure the redshift of the lenses and sources, probe their environments and model the mass potential of each lens system.

\cite{faure08a}, presented a sample of 67 new strong lens candidates discovered by visual inspection of bright/early type galaxies with photometric redshifts $<$1.0 in the COSMOS field. Because of the selection method, the sample potentially contains systems with exceptional individual properties. In this paper, we study one of these lens candidate systems: \5921 (RA=09$^{h}$59$^{min}$21.7$^{s}$, DEC=+02$^\circ$06\arcmin38\arcsec). The system consists of four point-like objects that lie on top of a perfect ring around an early-type galaxy. According to the best-fit spectral energy distribution, a photometric redshift of z$_l^{phot}$=0.45$\pm^{0.03}_{0.05}$, was inferred \citep{faure08a}. The morphology of the system suggests that it is a lensed AGN, the ring having been formed by the AGN's host galaxy, similar to that shown by RXSJ 1131-1231 \citep{sluse03} or PG 1115+080 \citep{impey98}.

We organize the paper as follows. In Sect. \ref{sec:dataset}, we present the large imaging and spectroscopic datasets. In Sect. 3, we analyze the nature of the system using information from all these datasets. The astrometry and photometry of the objects in the system obtained from light profile fitting procedures and techniques are presented in Sect. 4. In Sect. 5, we analyze the environment both around and along the line-of-sight to \5921. By using all the information gathered in the previous sections, in Sect. 6 we present mass models of the system, and in Sect. 7 we consider the different phenomena that can cause the observed flux ratio anomalies between the lensed images. Throughout the paper, we assume a flat cosmology with $\Omega_M$=0.3 and H$_0$=70 km s$^{-1}$Mpc$^{-1}$. All magnitudes are given in the AB system.

\section{Imaging and spectroscopic dataset}\label{sec:dataset}

\begin{figure}[h!]
	\centering
	\includegraphics[width=8cm,bb=0 0 779 390]{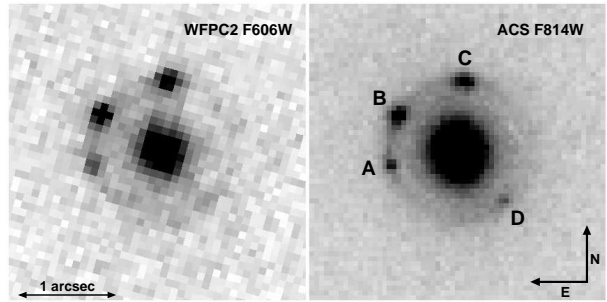}
	\caption{HST 3\arcsec\ side length cut-out images of \5921. The \textit{left panel} shows the WFPC2 F606W exposure and the \textit{right panel} shows the ACS F814W exposure with the naming scheme selected for the point-like objects.}
	\label{hst}
\end{figure}

\begin{figure}[h!]
	\centering
        \includegraphics[width=8cm,bb=0 0 702 1039]{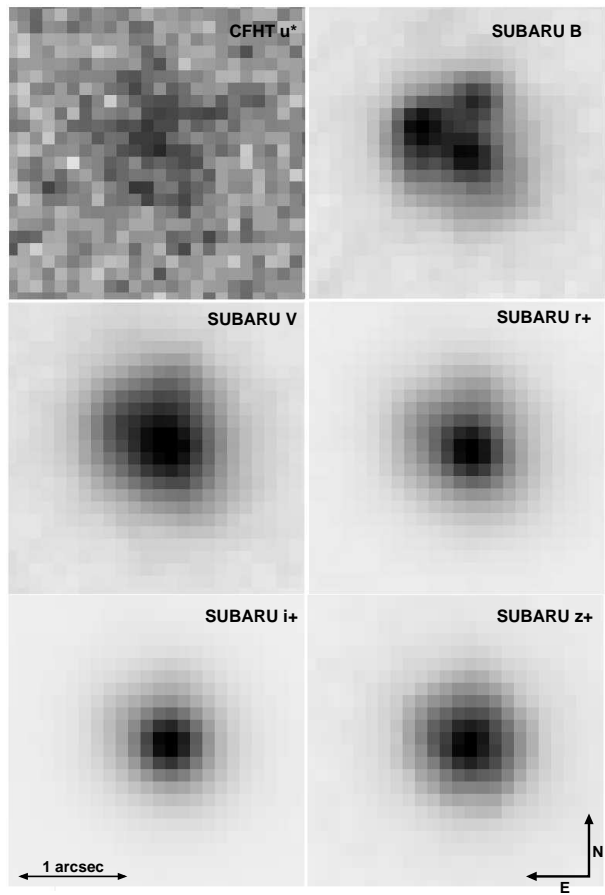}
	\caption{CFHT/Megacam and SUBARU/Suprime images of \5921. The 3\arcsec\ side length images are displayed from shorter (u$^\star$) to longer (z+) wavelengths \textit{from left to right, top to bottom}.}
	\label{allsub}
\end{figure}

\subsection{Imaging dataset from ground and space}
The COSMOS field \citep{scoville07} is a square field with 1.4 degrees side length ($\sim$2 square degrees). It was selected to be close to the celestial equator in a region with minimum extinction by dust in our galaxy ($<$$E_{(B-V)}$$>$$\backsimeq$ 0.02 mag), ensuring maximum observability and high observation depths. It has been observed using multi-band imaging from space and Earth. Thus, we have access to data of broad wavelength coverage for \5921\ and its neighborhood. Among these, the system was observed with the Subaru and CFHT ground based telescopes in the B, V, r+, i+, z+ and u$^\star$  bands \cite[data described in detail by][]{capak07}. From the Hubble Treasury programs \citep{scoville07}, we have HST/ACS observations in the F814W band \citep[the data reduction process is described in][]{koekemoer07}.  Additionally, we have access to WFPC2 F606W exposures of the system  (HST proposal id: 11289, PI: Kneib); part of the Strong Lensing Legacy Survey \citep[SL2S,][]{cabanac07}, where the system was also serendipitously found. In Fig. \ref{hst}, we display the HST exposures of the system with the chosen naming scheme for the point-like images (A to D, clockwise from the east-most image), which is used hereafter. In Fig. \ref{allsub} we show the CFHT and Subaru exposures. Even though the system was observed in both the radio \citep[1.4GHz VLA-COSMOS,][]{schinnerer07} and X-ray \citep[XMM/Newton-COSMOS,][]{brusa07}, no signal was measured \citep[as already mentioned in][]{faure08a}. We summarize the properties of the imaging dataset in Table \ref{tab:observations}.

\begin{figure*}[t]
	\centering
	\includegraphics[width=16cm,bb=0 0 1259 972]{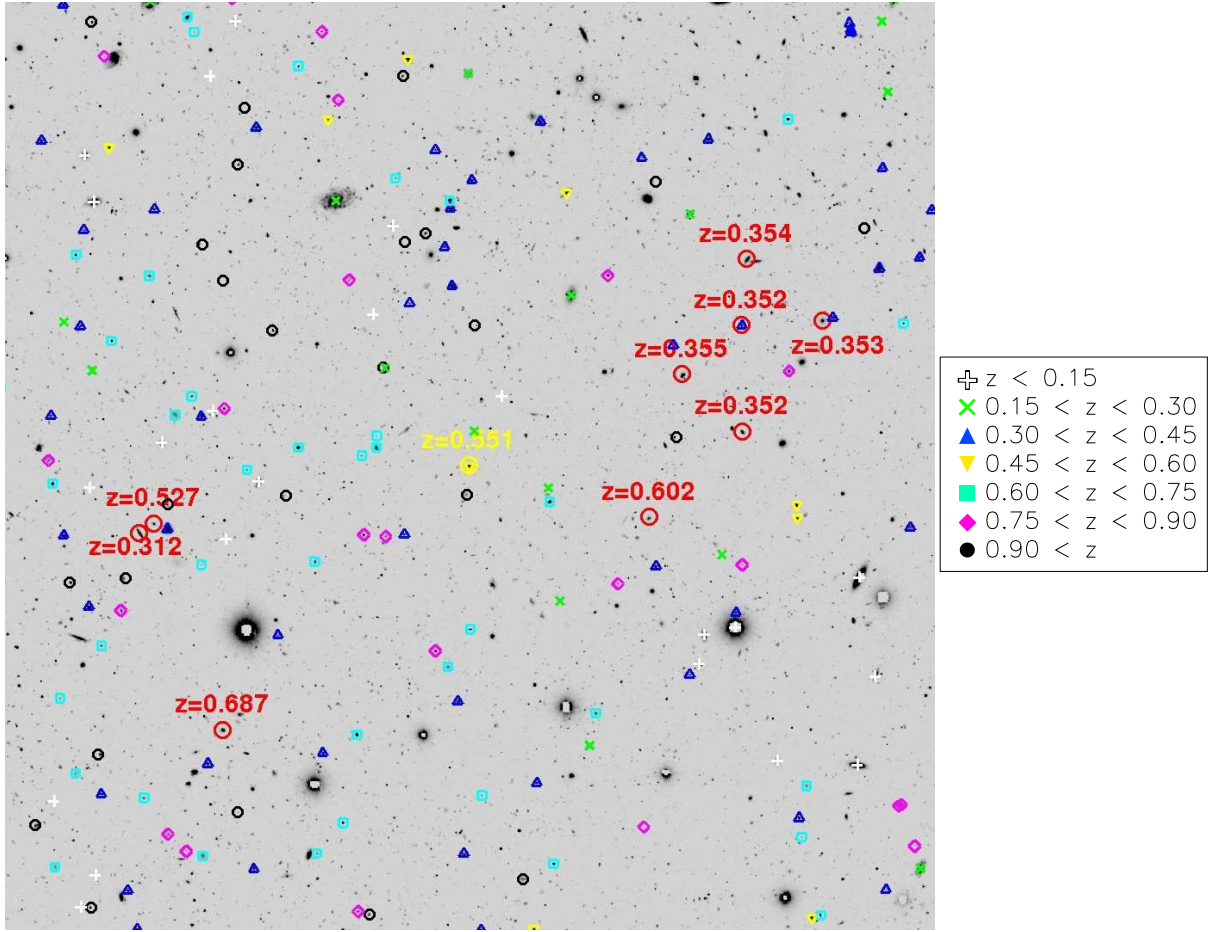}
	\caption{Subaru/Suprime B band image of the 8\arcmin$\times$8\arcmin field around \5921. The central lensing galaxy of the system is shown in the center. The galaxies with redshifts from the zCOSMOS catalog are displayed in symbol-coded redshift bins. Galaxies with measured redshifts from the FORS1 observations are labeled and shown with large circles.}
	\label{FORSz}
\end{figure*}

\begin{table}[!h] 
 \renewcommand{\arraystretch}{1.0}
 \centering
 \begin{center}
 \caption[Imaging data for \5921]{\label{tab:observations}Summary of the different imaging data used for this analysis. The magnitude limits for the COSMOS observations are those shown in \cite{capak07} for the ground-based observations and in \cite{koekemoer07} for the HST observations. (a) The u$^\star$ frame was created by combining observations taken between the years 2004 and 2005, thus, the total exposure time of the COSMOS field in this band varies. (b) 27.1 is the magnitude limit for a point-like object in the ACS exposure and increases in brightness to 26.1 for a 1\arcsec-wide extended object.}
 \begin{tabular}{c | c | c | c | c}
 \hline
    Camera & Band& Date &Exp. [s] & Limit [mag]\\
\hline
\hline
\multirow{5}{*}{Suprime}&B&2004-02-19 &4320 &27.3 \\
&V& 2004-02-18&3240 & 26.6\\
&r+&2004-01-19 & 2160& 26.8\\
&i+&2004-01-22 &2880 & 26.2\\
&z+&2004-01-21 &4320 & 25.2\\
\hline
Megacam &u$\star$ & 2004-2005$^{(a)}$& $\sim$40000 &26.4\\
\hline
ACS &F814w &2004-04-08 & 2028&27.1$^{(b)}$ \\
 \hline
WFPC2 &F606W &2008-01-07 & 1200& 26.5\\
\hline
\end{tabular}
 \end{center}
 \end{table}

\subsection{Spectroscopic dataset}

Galaxies in the COSMOS field have been spectroscopically observed as part of the zCOSMOS survey \citep{lilly07}. zCOSMOS is a large redshift survey within the COSMOS field using the VIMOS spectrograph installed at the VLT. The second data release of the survey (zCOSMOS-bright DR2, released in October 2008; \citealt{lilly08}), contains $\sim$10,000 galaxy spectra with associated redshifts. Galaxies in the field around \5921\ with zCOSMOS redshifts (from the DR2) are displayed in Fig. \ref{FORSz}.

\begin{figure}
	\centering
	\includegraphics[width=8cm,bb=0 0 726 725]{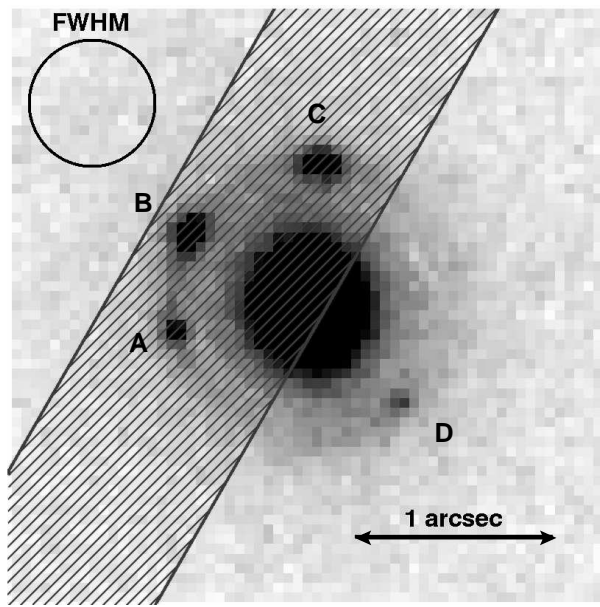}
	\caption{Placement of the slit (shaded region) containing three (brightest) images and the lensing galaxy in \5921 shown on top of a high resolution 3\arcsec\ side length ACS image cutout. The \textit{FWHM} of the VLT FORS1 observations is shown at the top left corner of the image. The coarse \textit{FWHM} plus the spectral dispersion direction (perpendicular to the orientation of the slit) ensures that the spectrum is highly contaminated by the much brighter galaxy.}
 	\label{slit}
 \end{figure}

Besides the zCOSMOS spectra and redshifts, the fields of 8 of the 67 strong lens candidates presented by \cite{faure08a}, including \5921, were observed with the FORS1 instrument at the VLT in Multi Object Spectroscopy (MOS) mode as part of a follow-up program (PI: Faure, Proposal ID: 077.A-0473(A)). The FORS1 observations around the field of \5921\ were obtained in April 2006 and are presented in this paper. They comprised a 7\arcmin$\times$7\arcmin\ field centered on \5921\ with the 150I grism (wavelength coverage: [3300 - 6500]\AA{}, resolution: 5.54\AA{}/pixel), with a total exposure time of 1800s. The standard CCD reduction and the spectra extraction was completed using pipeline recipes provided by ESO\footnote{ http://www.eso.org/sci/data-processing/software/pipelines/fors/fors-pipe-recipes .html}. The flux calibration was done using the long-slit spectrum of the LTT4816 standard star and \textsc{iraf}\footnote{\textsc{iraf} is distributed by the National Optical Astronomy Observatory, which is operated by the Association of Universities for Research in Astronomy (AURA) under cooperative agreement with the National Science Foundation.} routines. In Fig. \ref{slit}, we display the position of the central MOS slit, placed on top of the system: its goal was to measure the redshift of the central galaxy, as well as obtain signal from the close north-east point-like objects.

\section{The nature of the system}

The lens nature of an object is confirmed with the fulfillment of different steps, such as the corroboration of the appropriate number, relative brightness, and configuration of the candidate lensed images. The first step is, naturally, the confirmation of the line-of-sight alignment of the foreground ``lens'' and the background ``source'' with the observer by means of redshift measurements. However, in some cases, these can be difficult to confirm \citep[for an extended discussion see][]{marshall08}. For instance, the lens or the multiple images of a source can be too faint, making it difficult to acquire the spectroscopy required to measure their redshifts and corroborate that the candidate lens and source are foreground and background objects, respectively. This is precisely the case for \5921: the point-like images are very faint in comparison to the lensing galaxy (see Sect. \ref{sec:photom}). Nevertheless, an ensemble of evidence allow us to infer the gravitational lens nature of the system.

\begin{figure}
	\centering
 	\includegraphics[width=8cm,bb=40 0 510 284]{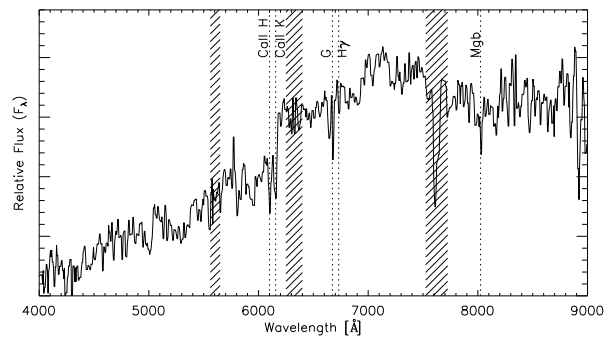}
	\caption{The flux-calibrated VLT/FORS1 spectrum of the central lensing galaxy towards \5921. Identified lines are labeled. Shaded areas denote regions of strong sky emission or absorption.}
	\label{galspec}
\end{figure}

\subsection{The multi-band images}
The distribution of the point-like images around the central galaxy shows a typical strong lens configuration: a pair of close images (A and B) and two images spread toward north (C) and south-west (D), similar to the configuration of the lensed quasars PG1115+080 \citep{schechter97}, MG0414+0534 \citep{hewitt92}, and WFI J2033-4723 \citep{morgan04}. Moreover, the point-like shape of the images suggests that the background source is an AGN. This hypothesis is also supported by the HST images, where a smooth structure forming a perfect Einstein ring below the point-like images, can be interpreted as emission from the host galaxy of the AGN. Based on this interpretation, we expect the point-like images to exhibit typical AGN emission lines such as Ly$\alpha$ ($\lambda$=1215\AA{}) or CIV ($\lambda$=1539\AA{}).

As shown in Fig. \ref{allsub}, the point-like structures have very similar wavelength dependence. In the CFHT/Megacam u$^\star$ band images (limiting magnitude=26.4, central wavelength=3797\AA{}), where there should be very low contamination by the central galaxy (based on the photometric redshift z$\sim$0.5, the 4000\AA\ break should be at $\sim$6000\AA), we are also unable to detect emission from the point-like images. The multiple images become visible in the Subaru/Suprime camera B band frames (limiting magnitude=27.4, central wavelength=4459\AA), where their brightness is comparable to that of the galaxy. Toward longer wavelengths, their light becomes rapidly contaminated by that of the galaxy. This behavior suggests that the point-like images are related to each other. Furthermore, it provides by itself an estimate of the redshift of the source, by a criterion similar to that present in the Lyman break technique \cite[e.g.,][]{giavalisco98} used to detect high-redshift star-forming galaxies. Assuming that our background source is an AGN, which are Ly$\alpha$ emitters, we can classify it as a u$^\star$ ``drop-out''. Thus, we can set the location of the Lyman break + Ly$\alpha$ line longwards of the u$^\star$ band central wavelength, which provides a lower limit to the redshift of the background AGN of z$\approx$3.

\subsection{The FORS1 spectra} 
\label{sec:FORS}

 From the slit covering a large portion of the system (see Fig. \ref{slit}), we obtain a single spectrum. Even though it contains emission from both the point-like objects and the central galaxy, the observed spectrum is consistent with a typical early type galaxy SED at a redshift of z=0.551$\pm$0.001 (see Fig. \ref{galspec}). This value was derived using the CaII H, CaII K, G, H$\gamma$, and Mgb absorption lines.

\begin{figure}[!ht]
	\centering
	\includegraphics[width=8cm,bb=40 0 504 360]{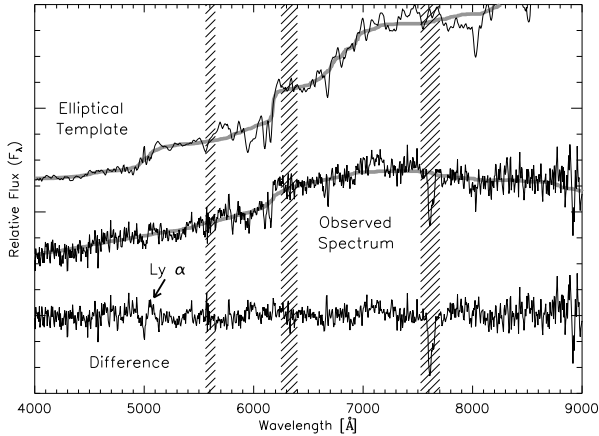}
	\caption{The \5921\ central galaxy spectrum. \textit{Top}: elliptical template spectrum. \textit{Middle}: measured spectrum of the lensing galaxy + point-like images. \textit{Bottom}: difference between the two spectra. The thick gray lines show the continua used to scale the respective spectra prior to subtraction. The difference spectrum shows an evident residual feature at around 5050 \AA{} that we interpret as Ly$\alpha$ at z=3.14. Shaded areas denote regions of strong sky emission or absorption.}
	\label{specsub}
\end{figure}

\begin{table}[!h]
\renewcommand{\arraystretch}{1.0}
\centering
\begin{center}
\caption{\label{forsztab} Galaxies with measured redshifts from the FORS1 observations. The measured errors for the redshifts are $\Delta$ z=10$^{-3}$.}
\begin{tabular}{c c c}

\hline
 RA & DEC &z\\
 \hline

09:59:09.69&02:07:53.6&0.353\\
09:59:12.21&02:08:25.4&0.354\\
09:59:12.39&02:06:56.4&0.352\\
09:59:12.41&02:07:51.2&0.352\\
09:59:14.44&02:07:26.0&0.355\\
09:59:15.59&02:06:12.0&0.602\\
09:59:21.79&02:06:38.7&0.551\\
09:59:30.27&02:04:22.0&0.687\\
09:59:32.69&02:06:08.8&0.527\\
09:59:33.24&02:05:47.6&0.312\\
\hline
\end{tabular}
\end{center}
\end{table}

Although the slit of the FORS1 observations on top of the system was aligned so as to contain three of the point-like objects (see Fig. \ref{slit}), at first sight no emission line from any of them is seen. The single spectrum that could be extracted is highly contaminated by the much brighter galaxy in the spectral range. However, when subtracting a scaled elliptical-galaxy template spectrum \citep{kinney96} from the observed spectrum (see Fig. \ref{specsub}), a small feature is observed at 5050\AA{}. We note, however, that the detection significance of this feature depends on the template selected for the subtraction.

The asymmetric shape of this feature suggests that it is Ly$\alpha$ emission. Furthermore, since the galaxy is brighter than the AGN, especially at longer wavelengths, and because of the photometric u$^\star$ drop-out criterion, we infer that Ly$\alpha$ is the only emission line that could be seen. The Ly$\alpha$ line is in general brighter than other broad emission lines and is located (in this case) at observed wavelengths corresponding to the fainter part of the galaxy's spectrum, thus, is less contaminated by the galaxy light. This line yields a redshift for the point-like images of z=3.14$\pm$0.05. This redshift is assigned by the single line and the error determined by its width.

From the spectra obtained from the remaining slits of the FORS1 observations, we derived the redshifts of 9 additional galaxies in the $\sim$7\arcmin\, (or $\sim$2.7 Mpc at z=0.55) field around \5921. In Fig. \ref{FORSz} and Table \ref{forsztab}, we report the location and redshift of these galaxies, 5 of which are at a similar redshift and relatively close to each other. This probably implies that they are members of a group or cluster of galaxies at z$\approx$0.35.

\subsection{Our interpretaion of the point-like objects}

If \5921\ is a genuine lensed AGN, as implied by the typical physical configuration of the system (four images of a source, a close pair, an Einstein ring formed by the host galaxy), the u$^\star$ drop-out criterion suggests that the redshift should be z$\gtrsim$3. The identification of a possible Ly$\alpha$ emission feature using the FORS1 spectrum is evidence that this is indeed the case and that the source is an AGN at a candidate redshift of z$_s$=3.14$\pm$0.05.

\section{Astrometry and photometry of the system}
\label{sec:photom}
A lensing mass model depends mostly on the relative positions of the source images and lensing galaxy. However, other parameters obtained from a quantitative measurement of the light distribution of the objects in the system (e.g., flux ratios, ellipticities, position angles, see for example \citealt{faure08c}) can also be useful as constraints or as a comparison. It is hence important to accurately measure the astrometry of the objects in the systems and take advantage of the additional information retrieved in the measurement process in order to develop and interpret a reliable mass model of the lens.

For this purpose, we applied the \textsc{galfit} software \citep{peng02} to the WFPC2 (F606W) and ACS (F814W) space-based observations. The software allows us to fit analytical two dimensional light profiles to the objects seen in a given CCD frame, in our case, the four point-like images of the background source and the lensing galaxy.

\subsection{The PSFs}
As an input, \textsc{galfit} requires the point spread function (PSF) of the observations to convolve it with the different analytical profiles. We obtained the PSF model for the WFPC2 observations using \textsc{tinytim}\footnote{http://www.stsci.edu/software/tinytim}. Because of our conclusion about the nature of the point-like images the PSF was created by assuming a power law continuum with $\alpha_\nu$=-0.46 \citep[fit from SDSS quasars;][]{vandenberk01} to avoid PSF-color artifacts. Since the WFPC2 frame of the system was created by combining 3 dithered exposures, the modeled PSF does not take into account convolution effects. Nevertheless, as shown in the next subsection, the WFPC2 data is not used to obtain any acute lensing constraints (e.g., astrometric parameters), but only the photometry of the objects, which is not significantly affected by convolution effects.

 The ACS PSF exhibits far greater temporal and spatial variations than the WFPC2 PSF. To obtain an accurate PSF model, we referred to \cite{rhodes06,rhodes07}, which contains a statistical study that yields a measure of the focus value $f$ (to a micro-metric accuracy) of the different COSMOS ACS exposures. Using this information for the ACS exposure of \5921\ ($f$$\sim$-4.5$\mu$m) we applied the \textsc{idl}/\textsc{tinytim} procedures presented in \cite{rhodes06}\footnote{http://www.astro.caltech.edu/$\sim$rjm/acs/PSF/} to generate PSF models, in particular choosing those in the location of the CCD where the system is located. The PSF was created by assuming a single monochromatic slice at $\lambda$=8140\AA{} (central wavelength of the F814W band).

\subsection{The light-profile fitting}

The lensing galaxy was fitted by a de Vaucouleurs profile \citep{vaucouleurs48}. The parameters that define this profile are: the effective radius (or half-light radius, $R_{eff}$), the central position, the ellipticity (defined as $e$=$\frac{a-b}{a}$; $a$: semi-major axis and $b$: semi-minor axis), the position angle (PA, defined in degrees measured from north to east), and the magnitude.

\begin{figure}
 \centering
 \includegraphics[width=8cm,bb=0 0 543 355]{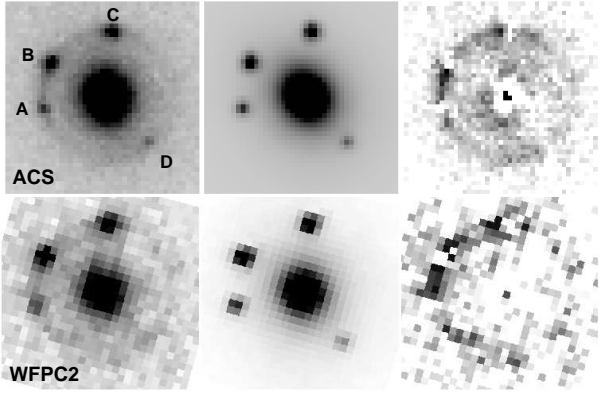}
 \caption{\textit{Top panel}: \textsc{galfit} de Vaucouleurs + 4 point-source fit to the ACS exposure of \5921: \textit{from left to right}: Original image, fitted model, and subtraction residuals. \textit{Bottom panel}: Fit of the WFPC2 exposures with the same display scheme.}
 \label{galfitim}
\end{figure}

\begin{table}[]
\renewcommand{\arraystretch}{1.0}
\centering
\begin{center}
\caption{\label{galfitgal} Parameters of the de Vaucouleurs fit to the lensing
galaxy towards \5921.}
\begin{tabular}{c c c c c}

\hline
   Filter &$R_{eff}$ [\arcsec]& e &PA &mag\\
\hline
F606W&0.45 (fix)&0.148(fix)&27.3 (fix)&21.72$\pm$0.02\\
F814W&0.45$\pm$0.01&0.148$\pm$0.009&27.3$\pm$2.0&20.310$\pm$0.009\\
\hline
\end{tabular}
\end{center}
\end{table}

The four point-like images were parameterized as point sources, which have central positions and magnitudes as free parameters. Because of the lower resolution of the WFPC2 observations, the parameters that define the de Vaucouleurs profile were fixed to those obtained with the ACS fitting (with the exception of the magnitudes and central positions). In both cases, the background was fixed to the median value of empty regions selected in the field. The fitted light profiles and residuals are displayed in Fig. \ref{galfitim}. The parameters obtained for the galaxy are summarized in Table \ref{galfitgal} and the photometry of the point-like images is presented in Table \ref{galfitqso}. Additionally, Table \ref{flratios} shows the flux ratios, with respect to the bright image B, derived from these magnitudes. The fitted positions of the different objects in the system are displayed in Table \ref{astgalfit}. Since the most accurate fit was obtained for the ACS exposure, we quote and use throughout the rest of this paper the astrometry obtained with this observation.

The error bars provided by \textsc{galfit} are based on the assumption that there has been a perfect fit to the data (i.e., $\chi^2$=1), which is not the general case. Hence, to obtain error bars for the different parameters of the fit, we created 500 Monte Carlo realizations for each of the HST frames and inferred the uncertainty from the scatter in the \textsc{galfit} results. These errors are in close agreement with those shown by a single \textsc{galfit} fit on the ACS data, but in the WFPC2 data, where the fit of the lensing galaxy is not as good, the errors delivered by \textsc{galfit} are slightly underestimated.

We attempted a light profile fitting of the ground-based observations. However, due mainly to the large PSF in these observations (\textit{FWHM}$\sim$1\arcsec) and the relatively small size of the system ($\sim$0.8\arcsec), the results were inconclusive. 

\begin{table}[]
\renewcommand{\arraystretch}{1.0}
\centering
\begin{center}
\caption{\label{galfitqso} Photometry of the lensed AGN images in \5921, magnitude values. Images labeled as in Fig. \ref{hst}.}
\begin{tabular}{c c c c c}
\hline
   Filter &A & B &C &D\\
\hline
F606W&25.60$\pm$0.08&24.57$\pm$0.04&24.83$\pm$0.06&26.50$\pm$0.18\\
F814W&25.42$\pm$0.06&24.59$\pm$0.03&24.53$\pm$0.03&26.19$\pm$0.10\\
\hline
\end{tabular}
\end{center}
\end{table}

\begin{table}[]
\renewcommand{\arraystretch}{1.0}
\centering
\begin{center}
\caption{\label{flratios} Flux ratios of the AGN images calculated with the magnitude values shown in Table \ref{galfitqso}.}
\begin{tabular}{c c c c c}
\hline
Date  & Filter &A/B & C/B &D/B\\
\hline
2004/04/08&F814W&0.47$\pm$0.03&1.06$\pm$0.04&0.23$\pm$0.02\\
2008/01/07&F606W&0.39$\pm$0.03&0.79$\pm$0.05&0.17$\pm$0.03\\
\hline
\end{tabular}
\end{center}
\end{table}

\begin{table}[]
\renewcommand{\arraystretch}{1.0}
\centering
\begin{center}
\caption{\label{astgalfit} Astrometry obtained with \textsc{galfit} for the background source images. All quoted values are in arcseconds with respect to the lensing galaxy at RA=09$^{h}$59$^{min}$21.768$^{s}$, DEC=+02$^\circ$06\arcmin38.33\arcsec}
\begin{tabular}{c c c c c c}
\hline
   Filter &&A & B &C &D\\
\hline
\hline
\multirow{4}{*}{F814W}&$\Delta_{RA}$&0.702&0.617&-0.067&-0.468\\
&&($\pm$0.005)&($\pm$0.003)&($\pm$0.003)&($\pm$0.008)\\
 &$\Delta_{DEC}$&-0.132&0.382&0.727&-0.507\\
&&($\pm$0.005)&($\pm$0.002)&($\pm$0.002)&($\pm$0.008)\\

\hline
\end{tabular}
\end{center}
\end{table}

\section{\label{sec:envi}The neighborhood of \5921}
Before developing a mass model of the lens, we considered the influence of secondary structures along the line-of-sight to the source that could perturb the properties of the lensing system. For this purpose, we use the zCOSMOS optical group catalog \citep{knobel09}. The catalog, created using spectroscopic information from the zCOSMOS redshift survey, contains 800 galaxy groups. They can be characterized by different quantities among which the ``fudge'' virial mass and the ``fudge'' virial velocity are the most useful for our purposes. The ``fudge'' quantities are obtained using the group richness and redshift as a means of obtaining the virial velocity and mass, adopting the observed relation between these and the mock catalogs created from numerical simulations for the COSMOS field. For a detailed discussion of the fudge virial mass we refer to \cite{knobel09} (the procedure to obtain the fudge virial velocity is analogous).

\begin{figure}
 \centering
 \includegraphics[width=8cm,bb=0 0 851 1415]{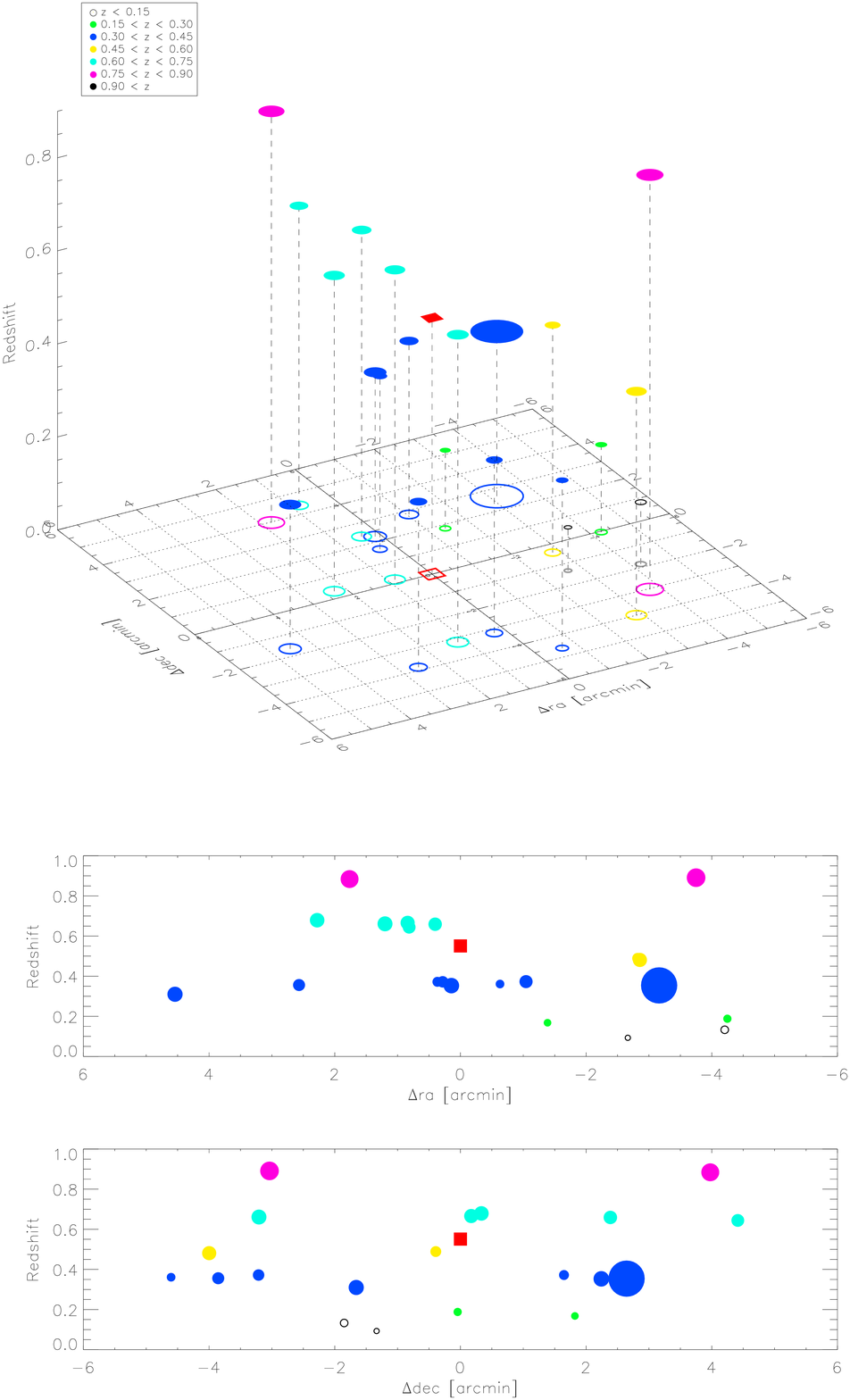}
 \caption{The 21 groups from the zCOSMOS optical group catalog around \5921, displayed as circles. The relative sizes of the circles scale with the ``fudge'' velocity dispersion. The lensing galaxy is displayed in the center as a square. In \textit{the top panel}, the groups are projected onto the observer's plane as empty circles for reference. \textit{The middle and bottom panels} additionally show the redshift distribution in right ascension and declination, respectively.}
 \label{grpproj}
\end{figure}

To study the gravitational influence of the optical groups on \5921, we selected the 21 groups located in a circle of $\sim$5\arcmin\ radius (or 2 Mpc at redshift z=0.551) centered on \5921 (see Fig. \ref{grpproj}). The location and redshift of the 5 galaxies at redshift z$\approx$0.35 obtained with the FORS1 observations (see Sect. \ref{sec:FORS}), coincides with the most massive group selected (ID: 20 in the original catalog). This massive galaxy group at redshift z=0.354 is the only one in our selection that has also been identified in the X-ray regime \citep{finoguenov07}.

Using the central position and mass (or velocity dispersion) of the 21 groups selected, we evaluate $\kappa$ and $\gamma$ at the position of the lens using three different assumptions for the mass profile of the groups: (i) point mass, (ii) singular isothermal sphere (SIS) and (iii) truncated isothermal sphere (TIS). To project the $\kappa$ and $\gamma$ values of the different foreground and background groups (obtained with each mass profile assumption) to the redshift of the lens, we used the methodology described in \cite{keeton03} \citep[see also,][]{momcheva06}. The point mass model assumes a singular total mass (which we selected as the virial mass), thus, there is no surface mass density outside the singularity and we can only observe the tidal shear $\gamma$ produced by this singularity. Assuming a singular isothermal sphere (SIS) profile for the groups, we can observe both the convergence and the shear (which are equivalent in this profile). Nevertheless, since the total mass of the SIS profile diverges, the values of $\kappa$ derived with it are certainly overestimated far from the center of each group. Truncating the profile eliminates this issue (TIS profile). The virial radius of each group was chosen for this truncation.

\begin{table}[h!]
\renewcommand{\arraystretch}{1.0}
\centering
\begin{center}
\caption{\label{tab:totalcont} Total contribution of the groups around \5921,
assuming the different profiles.}

\begin{tabular}{c | c | c | c}
\hline
&P. Mass&SIS&TIS\\
\hline
\hline
$\kappa_{tot}$&0.000&0.062&0.010\\
$\gamma_{tot}$&0.006&0.005&0.003\\
$\theta_{\gamma,tot}$&14$^\circ$&-43$^\circ$&-43$^\circ$\\
\hline
\end{tabular}
\end{center}
\end{table}

By adding the individual $\kappa$ and $\gamma$ values (or their vector sum in the case of the shear) of the 21 galaxy groups, we can determine the net influence of the environment (see Table \ref{tab:totalcont}). As expected, the total environmental convergence $\kappa$ produced by the groups, assuming a point mass potential, is zero. Assuming a SIS profile, we observe a total convergence $\kappa$=0.062, and for the TIS profile, a much smaller $\kappa$=0.010. From the latter, we can see that the total $\kappa$ contribution is negligible. However, it is important to remark that because the total convergence is a sum of positive scalar values, incompleteness in the catalog (i.e., unidentified or partially identified galaxy groups) leads to an underestimation of this value. On the other hand, the total external shear is a vector sum, thus, incompleteness in the catalog may lead to an under- or overestimation of the value. As a result, a smaller effect on the total sum of the shear would be observed (even though its direction can be compromised). Additionally, since shear is produced by tidal ``pulls'' from the mass distribution, the choice of potential profile does not have a large influence on the final result. As the total shear shown in Table \ref{tab:totalcont} is comparable to the same order of magnitude as the variance expected by cosmic shear, the $\gamma$ contribution exerted by the zCOSMOS groups in the catalog also cannot produce significant perturbations to the lens potential in \5921.

\section{The mass modeling of the lens}
To confirm the lens nature of the system and determine its mass distribution, we perform mass potential modeling of the lens by developing two types of mass models, grid-based and analytical. The grid-based models have the advantage of making no assumptions about the mass density profile of the lens. On the other hand, the analytical mass model, assumes a particular type of mass density profile, but provides information about the lensing galaxy that can be compared to observational quantities. In all models, we assume that the lens plane is at z$_l$=0.551 and that the source plane is at z$_s$=3.14.
 
\subsection{Grid-based mass model}

\begin{figure}
 \centering
 \includegraphics[width=8cm,bb=0 0 635 635]{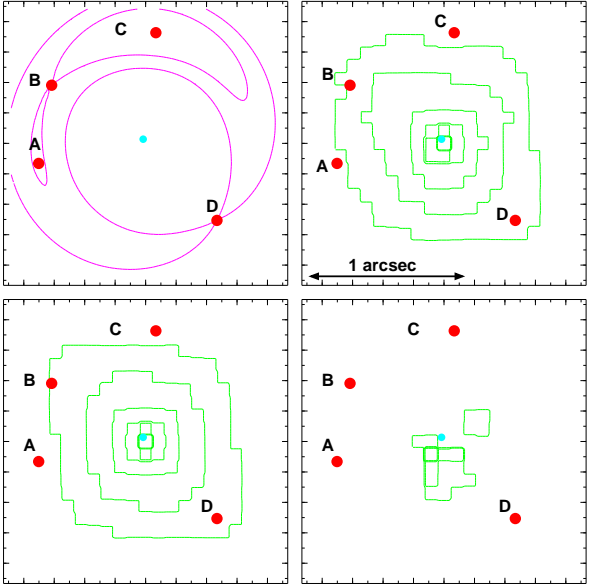}
 \caption{Grid-based model (including external shear) of \5921. \textit{The top left panel} shows the time-delay surface contours which map the location of the critical points (saddle, maxima and minima), and thus, the location of the images. \textit{The top right hand panel} shows the fitted mass distribution while \textit{the bottom left and right panels} show the symmetric and non-symmetric contributions to the mass distribution, respectively. Mass contours are defined as $-2.5log_{10}(M)$. The non-symmetric part accounts for approximately 7\% of the total mass.}
 \label{nonpar}
\end{figure}

The \textsc{pixelens} code \citep{saha04}\footnote{http://www.qgd.uzh.ch/projects/pixelens/} provides a grid-based mass model of the lens using the observed image positions as constraints. A requirement of the minimization is that the arrival time order of images is known. If incorrect image orders are entered, the model will either not converge or the fitted time-delay surface will determine critical points in locations not occupied by images of the background source. With this constraint, by trial and error, we can obtain the correct time delay order of the images (and the type of singularity on which they are placed), which in this case was: C(minimum)$\mapsto$A(minimum)$\mapsto$B(saddle)$\mapsto$D(saddle). Using this order, we obtain a pixelated profile for the lens. The time-delay contours for the correct image order are shown in Fig. \ref{nonpar} (top left panel).

	We tested two models, a first model without any external shear and a second model with external shear. In grid-based modeling, the mass profile is free and a non-symmetrical mass distribution is inferred from the fit. However, by dividing the mass distributions into symmetrical and non-symmetrical components (as shown in Fig. \ref{nonpar}, lower panel), we can see that the non-symmetric residual is rather small (for both models).

	For the model without an external shear, we obtain a mass profile with a slightly larger ellipticity than that of the observed light distribution of the galaxy. Additionally, the position angle of this mass distribution appears to be larger than 45$^\circ$ (compared to the 27.3$^\circ$ observed position angle of the light distribution). The amount of non-symmetrical mass in the mass profile accounts for approximately 10\% of the total mass.

	The model with an external shear (shown in Fig. \ref{nonpar}) shows an ellipticity similar to that of the observed light distribution with a position angle that is also similar to the observed one. In this case, the non-symmetrical residual of the mass distribution accounts for only 7\% of the total mass. It is therefore plausible that, even though small, the contribution of an external shear is required to model the system. We note that, although the contribution from the asymmetrical part of the mass profile located south west of the center of the lens galaxy is low, this may be indicative of some kind of substructure within the lens.

\begin{figure}
 \includegraphics[width=8cm,bb=0 0 504 360]{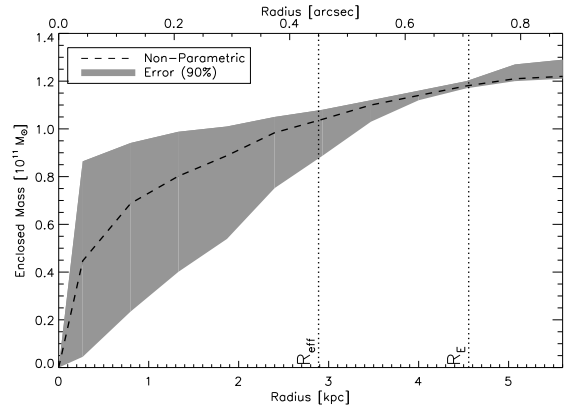}
 \caption{Radially enclosed mass obtained with the grid-based model that included external shear. The segmented lines display the effective radius fitted for the lensing galaxy and the measured Einstein radius of the system, respectively.}
 \label{slope}
\end{figure}

	 In both grid-based fits (with and without external shear), because of the steepness degeneracy \citep[e.g.,][]{falco85}, having as constraints only the position of the images (lying on a ring) and a low ellipticity, the slope of the profile is very degenerate (see Fig. \ref{slope}). Nevertheless, the total mass inside the Einstein radius (0.71\arcsec\ or 4.5 kpc at the redshift of the lensing galaxy) is well defined and equals $M_{R_E}$=1.2$\times$$10^{11} M_\odot$ in the cases with and without external shear.
	  
	Time-delays for both models are relatively similar but have large error bars. They are of the order of half a week for the longest delay (C to D) and of the order of a couple of hours for the shortest (A to B).

	Grid-based modeling \citep{saha04} provides insights on the mass distribution of the lens and a familiarization with the system. However, being grid-based, it naturally has enormous freedom in defining the mass profile and does not allow a quantitative comparison with observations. The final pixelated maps reflect the highest probability model from the ensemble of models tested (100 in our case).

\begin{table}[]
\renewcommand{\arraystretch}{1.0}
\centering
\begin{center}
\caption{\label{SIEtab} SIE fit to the lensing system. All observational constraints were fixed to the observed values. The galaxy's central position was allowed to move within 0.01\arcsec, which is the estimated uncertainty.}
\begin{tabular}{ c | c | c }
\hline
& SIE & SIE+$\gamma$\\
\hline
$\Delta$x [\arcsec]&0.01&-0.007$\pm$0.002\\
$\Delta$y [\arcsec]&0.005&-0.005$\pm$0.002\\
$\sigma$ [km/s]&190.7&189.3 $\pm$ 0.4\\
e&0.15 (fixed)&0.15 (fixed)\\
PA& 27.3$^{\circ}$ (fixed)& 27.3$^{\circ}$ (fixed)\\
$\gamma$&-&0.038 $\pm$ 0.002\\
 $PA_{\gamma}$&-&107.4 $\pm$ 1.3\\
$\chi_\nu^2$&$\sim$100&0.3\\

\hline

\end{tabular}
\end{center}
\end{table}

\subsection{Analytical mass model}

\begin{figure}
	\centering
	\includegraphics[width=8cm,bb=0 0 368 738]{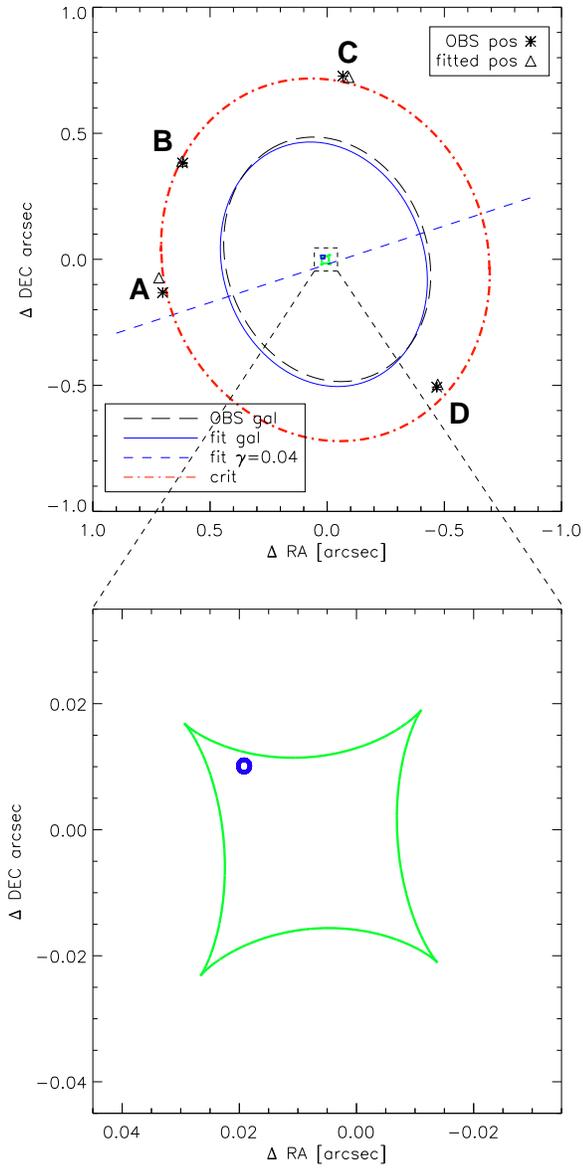}
	\caption{\textsc{lenstool}'s SIE+$\gamma$ fit to the lensing galaxy. Observed (asterisks) and resulting (triangles) image positions from the mapping of the position of the fitted background source (circle). The dashed-dotted curve shows the lens plane critical curve and the central continous diamond shaped curve shows the source plane caustics. The long-dashed and continous ellipses denote the observed and fitted lensing galaxy positions, respectively. The short-dashed line shows the orientation of the external shear required by the model. \textit{The bottom panel} shows a zoom to the center of the system.}
	\label{SIEfixmod}
\end{figure}

To obtain quantitative information about the mass model parameters and compare this with the information obtained by \textsc{pixelens}, we attempt to develop an analytical model of the system using \textsc{lenstool}\footnote{http://www.oamp.fr/cosmology/lenstool/} \citep{kneib93}.  \textsc{lenstool} does source and image plane parameter based $\chi^2$ minimization using a Bayesian algorithm (prior based minimization) and a Markov Chain Monte-Carlo (MCMC) process that samples the probability distribution by random variation of the parameters. This parameter space scanning, prevents the minimization being trapped in local $\chi^2$ minima and allows a very robust result with minor speed trade-off. The details of the algorithm and the underlying statistics are described in \cite{jullo07}. For our mass models, we apply the image plane minimization algorithm and use the observed positions of the AGN images as constraints, with a conservative error of $\Delta$=0.01\arcsec. This error accounts for imperfections in the lensing galaxy's position fit because of the overestimation of the flux in its central cusp.

	Constraining the galaxy's central position, ellipticity, and position angle to be the observed values and fitting a singular isothermal ellipsoid (SIE) yields a poor fit to the data ($\chi_\nu^2$$\approx$100), as expected from the results of the grid-based modeling.

When we add an external shear to our SIE model, we obtain far more accurate results (as displayed in Table \ref{SIEtab} and Fig. \ref{SIEfixmod}). Figure \ref{SIEfixmod} shows the comparison between the observed AGN image locations (asterisks) and the ``best fit'' image locations (triangles). These ``best fit'' image locations come from the mapping of the average location of the minimized source position. The zoom in the bottom panel, allows us to see the position of the source with respect to the caustics produced by the mass potential. The total mass within the Einstein radius matches the value inferred by the grid-based mass model ($M_{R_E}$$\approx$$1.2\times10^{11}M_\odot$).

\begin{figure}
 \centering
 \includegraphics[width=8cm,bb=0 0 459 350]{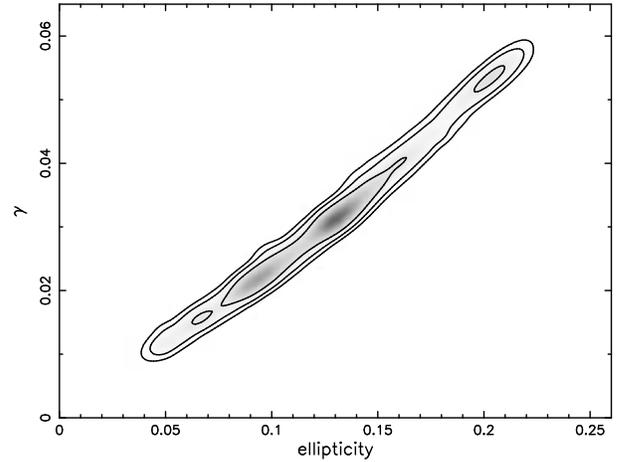}
 \caption{Degeneracy between the ellipticity of the lensing galaxy and the external shear. The contours show the confidence regions (68\%, 95\% and 99\%) for the fitted parameters.}
 \label{degenel}
\end{figure}

As seen in Table \ref{SIEtab}, the model needs to include a small external shear ($\gamma$=0.038) to reproduce the observed image configuration. The small strength of this shear is compatible with us not observing a strong influence from the environment. Nevertheless, its strength is somewhat larger than that implied by the group analysis. This may be indicative of: i) a misalignment between the light and the mass distribution of the lensing galaxy, ii) a contribution from individual galaxies close to the main lensing galaxy, and/or iii) a more complex description of the lens potential (e.g, non-isothermal slope or a misaligned dark matter halo). A misalignment between the light and mass distribution in this case is not a likely explanation since the model does not provide a good fit even when the ellipticity or position angle of the mass distribution are allowed to vary ($\pm$10$^\circ$ with respect to the position angle of the light distribution; see \citealt{keeton97}). We also examined the degeneracy between the ellipticity of the mass distribution and the external shear. To achieve this, we scanned the $\chi^2$ surface produced when these parameters are allowed to vary (in addition to the velocity dispersion, and central position), while the position angles of the mass profile and external shear are fixed to their observed and best-fit model values, respectively. Figure \ref{degenel} illustrates the clear degeneracy between these two parameters, as the necessary strength of the external shear decreases as the ellipticity of the mass profile decreases. However, the best-fit model value for the ellipticity (e$\approx$0.13) remains consistent with that measured from the light profile of the lensing galaxy (e=0.148), and a shear of strength $\gamma$$\gtrsim$0.01 is required at 99\% confidence. Regarding the contribution from individual galaxies, at the time being, there is no spectroscopic measurement of the immediate neighbors. Nevertheless, the three brightest and closest neighboring galaxies ($<$6\arcsec\ in projection to the observational plane) are aligned in an $\sim$105$^{\circ}$ east from north direction, in close agreement with the modeled external shear. Their photometric redshifts indicate that they are not within the same plane as the main lensing galaxy ($z_1$=1.1, $z_2$=0.8 and $z_3$=1.2). Assuming singular isothermal sphere profiles for these three galaxies and projecting their contribution (from their respective photometric redshifts) onto the lens plane (at $z_l$=0.551), they could reproduce the required external shear if they each have velocity dispersions of $\sigma$$\sim$300~km~s$^{-1}$. Even when velocity dispersions of that order have been measured in galaxies, the low brightness of these candidates ($m_{I,1}$=24.1, $m_{I,2}$=23.2 and $m_{I,3}$=24.5), makes it an unlikely possibility in these cases.

By considering an analytical model, different properties of the system can be checked both in the lens and the source plane. The fit of the potential can be used to measure the values of convergence ($\kappa$), shear ($\gamma$), and, thus, the magnification ($\mu$) expected at different positions in the lens plane. With these modeled magnification values at the position of the images, we can then infer the expected flux ratios for the different images. The $\kappa$, $\gamma$, and $\mu$ values and the flux ratios (with respect to image B) of the different images are displayed in Table \ref{tab:kapgam}. Comparing these modeled flux ratios with the observed ones (see Table \ref{flratios}), we can see a discrepancy. The possible explanations of this discrepancy are discussed in Sect. \ref{sec:anomalies}.

\begin{table}[]
\renewcommand{\arraystretch}{1.0}
\centering
\begin{center}
\caption{\label{tab:kapgam} Convergence ($\kappa$), shear ($\gamma$), and magnification ($\mu$) shown by the potential (SIE+$\gamma$) at the location of each lens plane image. The last column shows the flux ratio of each image with respect to image B inferred for the magnification values $\mu$.}
\begin{tabular}{ c | c | c | c | c}
\hline
Image&$\kappa$&$\gamma$&$\mu$&Fl. Ratio (/B)\\
\hline
A&0.472&0.508&47&0.70\\
B&0.509&0.506&67&1.00\\
C&0.498&0.469&31&0.46\\
D&0.548&0.525&14&0.21\\
\hline
\end{tabular}
\end{center}
\end{table}

The magnification values displayed in Table \ref{tab:kapgam} can also be coupled with the measured photometry of the individual point-like images (Table \ref{galfitqso}) to obtain an estimate of the intrinsic brightness of the background source. With this, we obtain an approximate absolute magnitude of the source of: $M_{606W}$$\sim$-17.5 and $M_{814W}$$\sim$-18.0. Even though these values are the median of the absolute magnitudes (with no k-correction) obtained for each image, which are affected by uncertainties related to flux ratio anomalies and the assumption of a SIE mass profile, they reflect the low luminosity nature of the source AGN \citep[LLAGN; e.g.,][]{storchi95,bower96,cid04}.

\subsection{Mass-to-light ratio}

As shown previously, theoretical mass modeling (both grid based and analytical) can provide an accurate estimate of the total mass enclosed within the Einstein radius. Using aperture photometry within this Einstein radius (i.e., $\theta_E$=0.71\arcsec), we also measured the amount of light enclosed within it, to compare it with the enclosed mass. This measurement was performed on the fitted \textsc{galfit} de Vaucouleurs profile, to avoid the contribution from lensed images.

We measured an ACS F814W aperture magnitude of 20.88 for the lensing galaxy. Using a k-correction of 0.45 for an elliptical galaxy at redshift z=0.55 in the F814W filter and an absolute solar magnitude of $M_{\odot_{F814W}}$=4.53\footnote{obtained from http://www.ucolick.org/$\sim$cnaw/sun.html}, we obtain $L_{F814W}$$\approx$$4.3\times10^{10}L_{\odot_{F814W}}$. This, together with the enclosed mass ($1.2\times10^{11}M_\odot$, see Sect. 6.1 and 6.2), leads to a mass-to-light ratio of $M/L_{F814W}$$\approx$$2.7M_\odot/L_{\odot_{F814W}}$. The WFPC2 F606W observations yield an aperture magnitude of 22.28 within the Einstein radius. Using a k-correction of 1.33 and an absolute solar magnitude of $M_{\odot_{F606W}}$=4.74, we obtain $L_{F606W}$$\approx$$3.3\times10^{10}L_{\odot_{F606W}}$, leading to a mass-to-light ratio of $M/L_{F606W}$$\approx$ $3.6M_\odot/L_{\odot_{F606W}}$.

\cite{fukugita95} show typical rest-frame elliptical galaxy colors of B-F814W$\approx$2.23 and B-F606W$\approx$1.29. We couple these colors with our aperture photometry (F814W and F606W) data and an absolute solar magnitude of $M_B$=5.33 to retrieve an estimate of the B-band luminosities for the lensing galaxy of $L_{B_{F814W}}$$\approx$$1.19\times10^{10}L_{\odot_{B}}$ using the F814W data, and $L_{B_{F606W}}$$\approx$$1.75\times10^{10}L_{\odot_{B}}$ using the F606W data). These lead, respectively, to: $M/L_{B_{F814W}}$$\approx$$10.1M_\odot/L_{\odot_{B}}$ and $M/L_{B_{F606W}}$$\approx$$6.9M_\odot/L_{\odot_{B}}$ which yield an average value of $M/L_B$$\approx$8.5$\pm$1.6. These values agree with those measured by \cite{keeton98} for the mass-to-light ratio within the Einstein radius of 10 elliptical lens galaxies ($M/L_B$$\approx$$10$ for galaxies at redshift z$\sim$0.55).

\section{Anomalous flux ratios}
\label{sec:anomalies}

From the observed flux ratios shown in Table \ref{flratios}, we can see that the ratios of the brightness of image A to B, and of D to B do not vary significantly (within the uncertainties) among the different filters and the different epochs (A/B 2004\_F814=0.47$\pm$0.03 and A/B 2008\_F606=0.39$\pm$0.03; D/B 2004\_F814=0.23$\pm$0.02 and D/B 2008\_F606=0.17$\pm$0.03). However, the C/B flux ratios show variations between different filters and/or over time (C/B 2004\_F814=1.06$\pm$0.04 and C/B 2008\_F606=0.79$\pm$0.05).

	Additionally, comparing the theoretical flux ratio values (Table \ref{tab:kapgam}: A/B=0.70, C/B=0.46, and D/B=0.21) with the observed flux ratios (Table \ref{flratios}), a disagreement can be seen. Even more generally, any symmetric model that we choose for the lens potential, expects images A and B to be the brightest images. They are the merging images of the lensing system and, consequently, the images with the highest magnifications. However, this is not the observed case, since image C is as bright as image B and brighter than image A in the ACS observations.

	Several phenomena can explain the flux differences between the lensed images from different bands and different epochs and their comparison with theoretical expectations, such as: galactic extinction, substructure, intrinsic variability of the background source, or microlensing \citep[e.g.,][]{anguita08b,yonehara08}:

\begin{itemize}

\item Galactic extinction: Galactic extinction is a static phenomenon (on human timescales) in which dust in the lensing galaxy absorbs the blue light originating in the background source images. This implies that lensed images located behind different column densities of dust appear to have color perturbations, in particular, those behind a larger dust column densities would appear redder. For this particular system, we obtain F606W-F814W colors of: 0.18, -0.02, 0.30, and 0.31 for images A, B, C, and D, respectively. In terms of the galactic extinction interpretation, there would be less dust accumulated over the projection of the lensing galaxy in the location of image B than in the others. Assuming galactic extinction \citep{cardelli89} with $R_v$=3.1, a differential column density of $\Delta N_H$$\approx$$1.8\times10^{21}$ would be required to explain the color difference between the anomalous images C and B ($\Delta_{F814W-F606W}$=0.32).

\item Mass substructures: In a similar way to galactic extinction, mass substructures can produce static anomalies in the measured flux ratios. A galaxy with mass substructures (of the order of $10^6$$\sim$$10^{10}M_\odot$), i.e., non-symmetrical luminous or dark matter components, or even unseen satellite galaxies, might modify the lensing potential, particularly the flux ratios \citep{kochanekdalal04}. In the case of \5921, even though it is small, the grid-based model of the lens shows a non-symmetric part of the distribution, which might be a hint of substructure in the lens. This phenomenon could explain, for example, the anomalously bright image C. However, because of the chromatic variation seen in the flux ratios, it cannot explain by itself the observed flux anomalies. If substructures and galactic extinction are responsible for the flux ratio anomalies, additional observations of the system should not display brightness and color variations with respect to the present data set.

\item Intrinsic variability: AGN vary their brightness over periods of years, months, weeks, and even days, usually becoming bluer as they become brighter. Unfortunately, the HST observations, although taken in different filters, are also taken at different epochs, which causes degeneracies in the interpretation of the chromatic effect. As the AGN images are separated between each other by time-delays (of the orders of hours to half a week, according to the mass modeling), the intrinsic variation of the AGN should be on timescales shorter than these to be able to see the anomalously bright image C, brighter than the close pair A and B. This is unlikely because quasars (statistically) require timescales of the order of years to produce brightness variations of $\sim$0.1 magnitudes \citep[for more details, see the structure function calculations of][]{vandenberk04}. Even if short timescale variability was occurring in this system, these rapid brightness fluctuations imply that it is unlikely after a period of 4 years, that slightly different flux ratios, although the same overall brightness order, are observed.

\item Microlensing: Microlensing is also a temporal phenomenon in ``macro''-lensed AGNs. In this case, stars in the lensing galaxy produce an additional (de)magnification of the ``macro''-lensed images of the source. Because of the projected location of the background source images in the lensing galaxy, microlensing is expected in this system. Much like intrinsic variability, (de)magnification by microlensing, in the general case, produces a blue excess: the blue emission from the background source originates in a smaller region, therefore, it is more effectively affected than the outer regions. As before, since the observations in different bands are taken at different epochs, no strong conclusions can be made from the chromatic point of view. The main difference with intrinsic variability/time-delay induced anomalies, is that in microlensing, there is no correlation between the multiple AGN images. Additionally, the projected velocities of the stars in the lensing galaxy, the lensing galaxy itself, and the background source set the microlensing timescales to be of the order of months to decades, compatible with the observed variations. For the microlensing interpretation, and considering that the measured flux in image C is higher than expected (with respect to image B), the flux anomaly that we observe is caused either by: (i) microlensing magnification of the flux from image C, mainly during the observations from 2004, (ii) de-magnification of image B in 2004, (iii) magnification of image B during the 2008 observations, or (iv) a combination of all phenomena. As image B is located on a saddle point (i.e., $\kappa+\gamma>1$), its de-magnification due to microlensing is more likely than for the other images \citep{schechter02}, which supports the second interpretation. In any case, since the four lensed images are located in regions of similar optical depth, variations in the brightness of all images should be seen in time.
 
\end{itemize}

\section{Summary and conclusions}

We have studied in detail the gravitationally lensed system candidate \5921\ discovered in the COSMOS field \citep{faure08a}. Using data derived from the COSMOS observational dataset plus additional HST-WPFC2 and VLT-FORS1 observations, we have obtained spectroscopic, astrometric, photometric, and morphological parameters of the objects in the field. The FORS1 observations allowed us to measure the redshifts of 10 galaxies in the field of \5921, including the central galaxy of the system at z=0.551$\pm$0.001. Coupling different criteria with the observations available, we conclude that this galaxy is lensing a background AGN and its host galaxy located at a tentative redshift z=3.14$\pm$0.05.

We have estimated the lensing contribution of the environment to the lensing system using the zCOSMOS optical group catalog. We have assigned different spherical mass profiles to the groups, constrained by their virial mass estimates. The analysis yields as a result a low environmental contribution at the lens position ($\kappa_{groups}$$\sim$0.01 and $\gamma_{groups}$$\sim$0.005). It is important to note that this analysis strongly depends on the completeness of the optical group catalog because the contribution from single undetected groups may influence the results. Furthermore, the individual group mass estimates in the catalog may be underestimated if the full mass structure of individual groups has not yet been identified. The forthcoming release of zCOSMOS 20K sample will allow us to address these issues in detail.

We have investigated the mass profile of the lensing system using strong lens modeling, both grid-based and analytical. Both types of modeling require a small but non-negligible external shear ($\gamma$=0.038$\pm$0.002) to reproduce the images' configuration. This presence of external shear cannot be fully explained by perturbations caused by either galaxy groups in the zCOSMOS optical group catalog or by realistic M/L ratio galaxies in the surroundings of the system. The requirement of this external shear may therefore be indicative of incompleteness in the environmental information we have at hand (i.e., incomplete identification of the structure of the groups), the need to consider more complex mass potentials (e.g., a non-isothermal slope or a misaligned dark matter halo), a contribution from undetected individual galaxies, or a combination of any of these possibilities. In a forthcoming paper (Faure et al. 2009), we investigate the dark matter halo alignment with the light profile for the rest of the COSMOS sample. In any case, for \5921, both measured and modeled external contributions to the potential are quite small; thus, the mass of the galaxy inside the Einstein radius has a low environmental contamination, making this system an interesting candidate for follow-up high resolution spectroscopical observations to measure the kinematics of the galaxy inside the Einstein radius.

The lens modeling of \5921\ shows time-delays of the order of hours to days between the different images and a total magnification of $\mu$$\approx$$150$. Comparing the observed brightness of the images with the magnification induced by the fitted mass profile infers the low luminosity nature of the background AGN with an intrinsic brightness of M$\sim$-17.5. Additionally, by comparing the lensing mass with the observed light enclosed within the Einstein radius ($\sim$4.5 kpc), we have inferred a mass-to-light ratio of the lensing galaxy of $M/L_B$$\approx$8.5$\pm$1.6.

Flux anomalies are observed between the different lensed images. With the available dataset, it is impossible to determine a unique reason for these variations. This is mainly because the observations available that can be used to measure accurate photometry for the point-like images are in different bands and at different epochs. This does not allow us to ascertain whether the phenomenon is dynamic or static. With this in mind, our preferred explanations of these incompatibilities are microlensing and/or substructures coupled with differential dust extinction in the lens, with microlensing in general being the most natural explanation of flux ratio anomalies in the optical range.

By completing follow-up observations of the system, either the phenomenon or phenomena responsible for these anomalies could be found and quantified. Multi-band photometry of the system at a single epoch would allow us to measure the color of the different images, free of possible temporal anomalies. In addition, by repeated multi-band observations at different epochs, the possible temporal flux variations could be proven and quantified. If active microlensing is present in this system, the additional resolution power brought by microlensing could infer interesting properties about the background source. If the anomaly is static, a measure of possible substructures and projected dust densities in this lens can be accomplished.

The lensing-induced magnification of the faint high-redshift background AGN in \5921 allows us to study the properties of an object that would have otherwise remained undetected in the COSMOS field. Even though low luminosity AGN (LLAGN) are expected to be far more common than high luminosity AGN, only a few of these objects have been observed beyond the local universe.

The small separations and the low brightness of the images of the background AGN in \5921 make it an ideal candidate for additional space-based and/or new generation optics ground-based follow-up observations to address the remaining open issues and understand the nature of the background AGN.

\begin{acknowledgements}
The authors would like to thank Raphael Gavazzi (IAP) for providing the data reduction of the WFPC2 dataset. TA acknowledges support from the International Max Planck Research School for Astronomy and Cosmic Physics at the University of Heidelberg. JPK acknowledges support from CNRS, CNES and the ANR grant ANR-06-BLAN-0067. ML acknowledges the Centre National d'Etudes Spatiales (CNES) for their support. The Dark Cosmology Centre is funded by the Danish National Research Foundation.
\end{acknowledgements}
\bibliographystyle{aa} 
\bibliography{12091}

\end{document}